\documentclass[twocolumn]{jpsj2}

\def\gsim{\mathop {\vtop {\ialign {##\crcr 
$\hfil \displaystyle {>}\hfil $\crcr \noalign {\kern1pt \nointerlineskip
 } 
$\,\sim$ \crcr \noalign {\kern1pt}}}}\limits}
\def\lsim{\mathop {\vtop {\ialign {##\crcr 
$\hfil \displaystyle {<}\hfil $\crcr \noalign {\kern1pt \nointerlineskip
 } 
$\,\,\sim$ \crcr \noalign {\kern1pt}}}}\limits}

\def\runauthor{Hiroyasu Matsuura et al,
}
\title{Theory of Mechanism of $\pi$-d Interaction in Iron-Phthalocyanine 
}          
\author{ Hiroyasu {\sc Matsuura}$^{1}$, Masao {\sc Ogata}$^{1}$, Kazumasa {\sc Miyake}$^{2}$, and Hidetoshi {\sc Fukuyama}$^{3}$
}
\inst
{
$^{1}$Department of Physics, University of Tokyo, 7-3-1 Hongo, Bunkyo, Tokyo 113-0033, Japan\\
$^{2}$Division of Materials Physics, Department of Materials Engineering Science \\
Graduate School of Engineering Science, Osaka University, Toyonaka, 
Osaka 560-8531, Japan\\
$^{3}$Faculty of Science and Research Institute for Science and Technology (RIST)
Tokyo Univerisity of Science\\ 
1-3, Kagurazaka, Shinjuku-ku, Tokyo 162-8601, Japan
}

\recdate
{
\today
}

\abst
{
Transition metal-phtahalocyanine(Pc) compound, TPP[Fe(Pc)(CN)$_2$]$_2$, which
is one of molecular conductors of charge transfer type with 3/4-filled
conduction band consisting of LUMO of Pc molecules, shows characteristic
features in transport and magnetic properties resulting from localized
magnetic moments $S=1/2$ associated with Fe$^{+3}$ atoms. 
We construct an effective tight-binding model of this system and study the mechanism of exchange interaction, $J$, between $d$ and $\pi$ electrons based on both second order perturbation of transfer integrals between $d$ and $\pi$ orbitals and numerical diagonalization.
It is found that there is no hybridization between $d$-orbitals and LUMO of $\pi$-orbitals and then super-exchange interaction in the Anderson model does not exist.
Instead, processes associated with Hund's rule both on $d$ and $\pi$ orbitals, which may be called "the double Hund's exchange mechanism", turn out to play important roles and the sign of resultant $J$ can be either ferromagnetic or antiferromagnetic depending on model parameters because of competition among various processes. 
By taking account of magnetic anisotropy due to spin-orbit interactions and comparing with experimental results, it is indicated that $J$ is antiferromagnetic and of the order of 100K.
}
\kword{Phtharocyanine (Pc), TPP[Fe(CN$)_2$(Pc)$]_2$, PNP[Fe(CN$)_2$(Pc)], TPP[Co$_{1-x}$Fe$_{x}$(CN$)_2$(Pc)$]_2$, Anisotropic g-values, $\pi$-d interaction 
}
\begin{document}
\sloppy

\maketitle

\section{Introduction}
Phthalocyanine (Pc) compounds with transition metal (TM) are organic materials with a variety of geometry together with the interplay between conduction electrons and magnetic moments of TMs.
They show many interesting phenomena such as charge ordering, giant magnetoresistance, Kondo effect, ferrimagnetism, and so on.

Recently, it has been reported that one of the TM-Pc systems, TPP[Fe(Pc)(CN)$_2$]$_2$, shows large negative magnetoresistance~\cite{Inabe1,Hanasaki1,Hanasaki2}.
Because the valence values are TPP$^{1+}$, CN$^{1-}$, Pc$^{3/2-}$ and Fe$^{3+}$ (3d$^5$ and the low spin state), this compound has a one-dimensional conduction band of 3/4 filling constructed from a lowest unoccupied molecular orbital (LUMO) of Pc, and has local moments of $S=1/2$ coming from the d orbitals of Fe.
On the other hand, a similar compound, TPP[Co(Pc)(CN)$_2$]$_2$, shows large positive magnetoresistance~\cite{Hanasaki3}. 
This system has a similar one-dimensional conduction band as in Fe-Pc compound, while it does not have the local moment on Co ($S=0$).
It is also known that TPP[Fe$_{0.07}$Co$_{0.93}$(Pc)(CN)$_2$]$_2$ shows negative magnetoresistance as in Fe-Pc~\cite{Hanasaki4}.
From these results, it is natural to think that the origin of the negative magnetoresistance is due to the interaction between the conduction electrons ($\pi$ orbitals on Pc) and the local moment on Fe.
We call this $\pi$-d interaction.

Assuming that the $\pi$-d interaction is antiferromagnetic (AFM), the origin of negative magnetoresistance was discussed on the basis of an extended Kondo model~\cite{Hotta1}.
On the other hand, recently it was claimed that the $\pi$-d interaction is ferromagnetic (FM) on the basis of quantum chemical calculation~\cite{Yu}.
On the basis of the ferromagnetic $\pi$-d interaction, the electronic state of the extended Kondo model has been discussed~\cite{Otsuka,Hotta2}.
Then, it is suggested that the origin of the negative magnetoresistance is double exchange interaction between Fe and Fe~\cite{Kimata}.
However, the negative magnetoresistance has been observed in TPP[Fe$_{0.07}$Co$_{0.93}$(Pc)(CN)$_2$]$_2$ as introduced above, and the result suggests that the origin of the negative magnetoresistance is not the double exchange mechanism, because the mean distance between Fe and Fe is about 70$\AA$.

Based on these situations, it is important to study the $\pi$-d interaction microscopically.
Although the amplitude and its sign of the $\pi$-d interaction was evaluated by the quantum chemical calculations, it is suspected that the quantum chemical calculations only take account of the low energy states. 
Thus, we think that the sign and the amplitude of the interaction is not decisive in real compounds. 
The mechanism of $\pi$-d interaction itself also has been unclear in the quantum chemical calculation.

In this paper, we study the electronic states of [Fe(Pc)(CN)$_2$] in detail, and clarify the mechanism of $\pi$-d interaction. 
First, by analyzing the electronic state of Pc, we find that there is no hybridization between d-orbitals and LUMO of $\pi$-electrons.
This means that there is no super-exchange interaction between d electrons and $\pi$ electrons.
Instead, the mechanisms of exchange interaction are "generalized" Kanamori-Goodenough mechanisms.
In particular, we find that a second order process which uses two kinds of Hund's rule couplings both on the d and $\pi$ electrons plays an important role.
This mechanism gives an antiferromagnetic exchange interaction even if it is not a super-exchange interaction.
We call this "double Hund's exchange" mechanism.

We find that the sign of the $\pi$-d exchange interaction can be both ferromagnetic and antiferromagnetic depending on the model parameters because of the presence of the competing processes. 
However, we think that the sign of the $\pi$-d interaction can be determined by a comparison with experiments.
When we discuss the comparison with experiments, it is necessary to take account of the anisotropy of the $\pi$-d interaction.
For this purpose, we consider a material, PNP[FePc(CN)$_{2}$], which has the same unit FePc(CN)$_2$ as in TPP[FePc(CN)$_{2}$]$_2$~\cite{Hanasaki1}.
The unit of FePc(CN)$_2$ has a local magnetic moment of $S=1/2$ of Fe atom, while it has no $\pi$ conduction electrons, because the valence is PNP$^{1+}$, CN$^{1-}$, Pc$^{2-}$ and Fe$^{3+}$.
In this case, we can study the electronic state purely of d electrons on Fe site without $\pi$ electrons.  
Therefore we discuss the electronic state of d electrons in this PNP compound by analyzing the anisotropy of the g values observed in EPR experiment.
On the basis of these results, we derive the effective model of $\pi$-d interaction, and discuss its sign by comparing with susceptibility obtained in the dilute system of Fe atoms, TPP[Co$_{0.93}$Fe$_{0.07}$Pc(CN)$_2$]$_2$.

This paper is organized as follows. In \S 2, the electronic state of Pc is discussed in the tight binding approximation based on $\pi$ orbitals on atoms in Pc.
We show that the hybridization between LUMO and d orbitals will varnish in this system.
In \S 3, on the basis of the electronic states of $\pi$ orbitals clarified in \S 2, the exchange interaction between d electrons and $\pi$ electrons is derived and the amplitude of this interaction is estimated. 
In \S 4, to consider the anisotropy of $\pi$-d interaction, the electronic state of d electrons is studied by analyzing the g factor in PNP[FePc(CN)$_{2}$] on the basis of a simple model including a crystalline electric field (CEF) and a spin-orbit coupling.
In \S 5, the effective model of $\pi$-d interaction is derived by using the results of \S 3 and \S 4, and the sign of the $\pi$-d interaction is discussed by the comparison with experimental results.

\section{Electronic state of Phthalocyanine}
\begin{figure}[h]
\begin{center}
\rotatebox{0}{\includegraphics[angle=-90,width=1\linewidth]{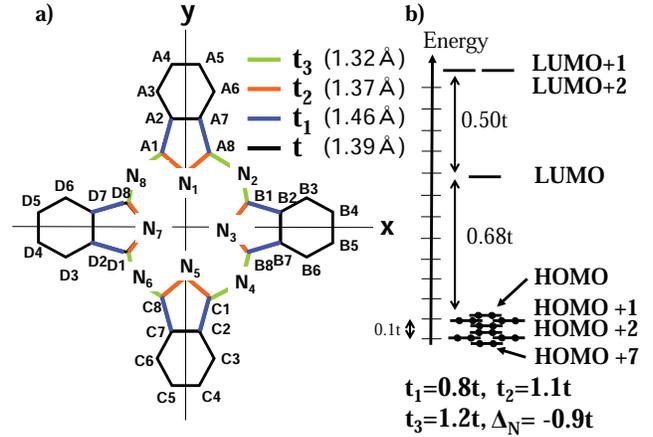}}
\caption{(Color online) (a) Structure of metal-free phthalocyanine. 
N$_{i}$ ($i$=1 $\sim$ 8) is $i$-th nitrogen (N) site, and A$_{1}$ $\sim$ D$_{8}$ are carbon (C) sites.  
$t$(black line), and $t_{1}$(blue line) are the transfer integrals between C atoms in hexagons, those between C atoms in pentagons, respectively.
$t_{2}$(red line), and $t_{3}$(green line) are those between C and N. 
(b) Energy level scheme of $\pi$ orbitals.
We also introduce $\Delta_{N}$ which is the one-body level of N measured from that of C.
  }
\label{Fig1}
\end{center}
\end{figure}

Figure\ \ref{Fig1}(a) shows the structure of metal-free phthalocyanine (Pc).
Here N$_{i}$ ($i$=1 $\sim$ 8) represents the $i$-th nitrogen (N) site, and A$_{1}$ $\sim$ D$_8$ are carbon (C) sites, respectively.  
The electronic state of Pc has been understood in detail by both experiments (XAS and XPS) and quantum chemical calculations~\cite{Chen,Edwards,Schaffer,Henriksson,Orti1,Orti2}.

In the following, we show that these experimental and calculational results can be semiquantitatively reproduced by the tight binding approximation using $\pi$ orbitals.
As defined in Fig. \ref{Fig1}(a), the transfer integrals $t$(black line), and $t_{1}$(blue line) are the transfer integrals between C atoms in hexagons, those between C atoms in pentagons, respectively.
$t_{2}$(red line), and $t_{3}$(green line) are those between C and N. 
We diagonalize this tight-binding model and adjust the parameters, $t \sim t_{3}$ so as to reproduce the results of the quantum chemical calculations.

Figure\ \ref{Fig1}(b) shows the energy levels of the $\pi$ orbitals when we choose $t_{1}= 0.8t$, $t_{2}=1.1t$, $t_{3}=1.2t$, and $\Delta_{N}=-0.9t$. 
This result is consistent with the quantum chemical calculations quite well. 
Here, $\Delta_{N}$ is the one-body level of N measured from that of C.

As shown in Fig.\ \ref{Fig1}(b), the gap between LUMO and LUMO+1, +2 is of the order of 0.5$t$.
Since $t \sim 3$eV, this gap is estimated as $1.5$ eV, which is in good agreement with the quantum chemical calculations (1.8eV for $H_2$Pc molecule, and 1.6 eV for LiPc)~\cite{Edwards,Orti2}.
It is noted that the energy difference between the highest occupied molecular orbital (HOMO) and HOMO+7 is tiny ( 0.1$t$ $\simeq$ 0.3eV).
Indeed this difference is estimated as 0.14 eV for LiPc in the quantum chemical calculations~\cite{Orti2}.

The wave function of the $i$-th $\pi$ orbital is expressed as follows:
\begin{eqnarray}
\Phi_{i}=\sum_{j=1 \sim 8}c_{j}^i\phi_{N_j}+\sum_{j=A_1 \sim D_8}c_{j}^i\phi_{j},
\end{eqnarray}
where $\phi_{N_j}$ and $\phi_{j}$ are the wave functions of the $\pi$ orbitals of $j$-th N and C, respectively.
The coefficient, $c_{j}^i$, is the amplitude of the wave function of $j$-th N or C atom in the $i$-th $\pi$ orbital.
Some of them ($c_{N_j}^i$, $j$ =1,3,5,7) are shown in Table \ref{Table1}.
\begin{table}[htbp]
\begin{tabular}{cc|c|c|c|c}
\hline
& \    i  & $c_{N_1}^i$ & $c_{N_3}^i$& $c_{N_5}^i$ & $c_{N_7}^i$  \\ \hline \hline
LUMO+1&$\Phi_{1}$ & 0.19 &-0.12   &-0.19  & 0.12 \\
LUMO+2&$\Phi_{2}$ & 0.12 & 0.19   &-0.12  &-0.19  \\
LUMO&$\Phi_{3}$   & 0    & 0      & 0     & 0    \\
HOMO&$\Phi_{4}$   & 0.37 & 0.37   & 0.37  & 0.37  \\
HOMO+1&$\Phi_{5}$ & 0.22 &-0.18   &-0.22  & 0.18  \\
HOMO+2&$\Phi_{6}$ &-0.18 & -0.22  & 0.18  & 0.22   \\
\hline
\end{tabular} 
\caption{The amplitude of coefficients of $\pi$ orbitals for the four central N atoms, i.e, N$_{j}$(j=1,3,5,7) sites.
}
\label{Table1}
\end{table}
These coefficients of N sites on LUMO are zero, while those on LUMO+1, LUMO+2, and HOMO, HOMO+1 and HOMO+2 are finite.
As discussed later, these coefficients are important to discuss the hybridization between the d orbitals in TM and $\pi$ orbitals, and we find that there are no hybridization between d orbital and LUMO.
In addition, we neglect HOMO+3 $\sim$ HOMO+7 in Table \ref{Table1}, because the hybridizations between these orbitals and  d orbitals are much smaller than the hybridization between LUMO+1 and d orbital as shown in \S 3.

\begin{figure}[h]
\begin{center}
\rotatebox{0}{\includegraphics[angle=-90,width=1\linewidth]{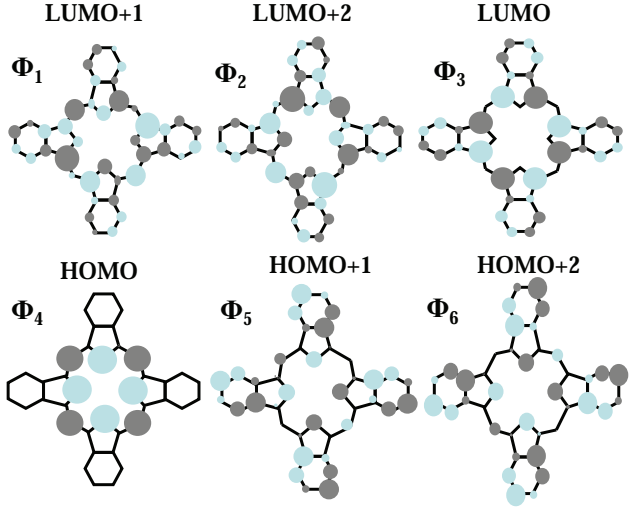}}
\caption{(Color online) Schematic pictures of wave functions of LUMO+1($\Phi_{1}$), LUMO+2($\Phi_{2}$), LUMO($\Phi_{3}$), HOMO($\Phi_{4}$), HOMO+1($\Phi_{5}$), and HOMO+2($\Phi_{6}$), respectively.
}
\label{S1_Fig2}
\end{center}
\end{figure}

The schematic pictures of wave functions are illustrated in Fig. \ref{S1_Fig2}.
The size of circle on each atom corresponds to the amplitude of the coefficient of the wave function, and the blue (black) circle indicates positive (negative) sign. 
The composition of these wave functions is consistent with the quantum chemical calculations.

\section{Mechanism and estimation of the exchange interaction between the d electron in Fe and $\pi$ electron in Pc}
\begin{figure}[h]
\begin{center}
\rotatebox{0}{\includegraphics[angle=-90,width=1\linewidth]{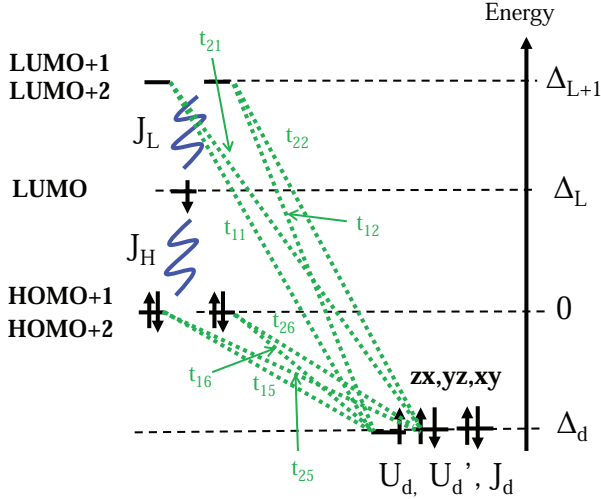}}
\caption{(Color online) Schematic picture of effective model.
The green dashed lines indicate the transfer integrals between d orbitals and $\pi$ orbitals.
We neglect HOMO, HOMO+3, HOMO+4, HOMO+5, HOMO+6, HOMO+7, $d_{3z^2-r^2}$, and $d_{x^2-y^2}$.
}
\label{Fig4a}
\end{center}
\end{figure}

In this section, we discuss the $\pi$-d interaction in TPP[FePc(CN)$_{2}$]$_{2}$ using the wave functions of $\pi$-orbitals in Pc obtained in the previous section. 
Although in TPP[FePc(CN)$_2$]$_2$ the nominal values are Fe$^{3+}$ and Pc$^{-3/2}$, we discuss the simple case of Pc$^{-}$ and Fe$^{3+}$, i.e,  there is one electron in LUMO of Fig. \ref{Fig1} and five electrons in 3d orbitals of Fe atom (see Fig. \ref{Fig4a}). 
As a microscopic model, we consider the following cluster model Hamiltonian constructed from the d orbitals of Fe and $\pi$ orbitals of Pc;
\begin{eqnarray}
H =H_{t} + H_{d} + H_{Pc},  \label{model_Mec}
\end{eqnarray}
where $H_{t}$ contains the transfer integrals between the d and $\pi$ orbitals, and $H_{d}$ and $H_{Pc}$ are Hamiltonians for the d and $\pi$ orbitals, respectively.

First, $H_{t}$ in eq. (\ref{model_Mec}) is given as 
\begin{eqnarray}
H_{t} = -\sum_{i=1}^3\sum_{j=1 (j \neq 4)}^6 t_{ij}(d_{i\sigma}^{\dag}\Phi_{j\sigma}+ h.c.), \label{hopping}
\end{eqnarray}
where $t_{ij}$ is the transfer integral between d$_{i}$ orbital and $\Phi_{j}$ orbital, where we define $d_{1\sigma} \equiv d_{yz\sigma}$, $d_{2\sigma} \equiv d_{zx\sigma}$, and $d_{3\sigma} \equiv d_{xy\sigma}$, and $\Phi_{1}=$ LUMO+1, $\Phi_{2}=$ LUMO+2, $\Phi_{3}=$ LUMO, $\Phi_{5}=$ HOMO+1, and $\Phi_{6}=$ HOMO+2 in Fig. \ref{S1_Fig2}, respectively.
We have neglected the other $\pi$ orbitals, because their energy levels are much higher (or lower) than that of LUMO, and the hybridizations with d orbitals are much smaller than those of LUMO+1.

There is no hybridization between LUMO (conduction band) and d orbitals (localized spins).
Therefore, the mechanism of $\pi$-d interaction is different from the mechanism based on the super-exchange interaction as in the impurity Anderson model which uses hybridizations~\cite{Anderson}.
Thus, we consider different mechanisms for the $\pi$-d interaction in the Fe-Pc complex.

In eq. (\ref{hopping}), we estimate the transfer integrals as $-t_{11}=t_{22}=0.38t_{dN}$, $t_{12}=t_{21}=0.24t_{dN}$, and $t_{15}=t_{26}=-0.44t_{dN}$, and $-t_{16}=t_{25}=0.36t_{dN}$. 
Here, $t_{dN}$ is the transfer integral between $d_{zx}$ (or $d_{yz}$) orbital and $p_{z}$ ($\pi$) orbital of the nearest-neighbor N atoms, and the coefficients of $t_{ij}$ have been determined from the amplitude of the wave functions of $\pi$-orbital $c_{j}^i$ in eq. (1) and from the symmetry of the t$_{2g}$ orbitals. 
It is important in the following considerations that $d_{xy\sigma}$ ($d_{3\sigma}$) orbital does not hybridize with any $\pi$ orbitals due to the symmetry requirement: i.e., $t_{3j}=0$.
Non zero transfer integrals are shown in Fig. \ref{Fig4a} by the green dashed lines. 
 
Next, $H_{d}$ in eq. (\ref{model_Mec}) is given as follows:
\begin{eqnarray}
H_{d} &=& \sum_{i=1 \sim 3,\sigma}\Delta_{i}^d n_{i\sigma} + U_d \sum_{i=1 \sim 3}n_{i\uparrow}n_{i\downarrow} \\
      & &  +\frac{U_d^{\prime} - J_d}{2} \sum_{i,j=1 \sim 3(i \neq j)} \sum_{\sigma} n_{i\sigma}n_{j\sigma} \\ 
      & & +\frac{U_d^\prime}{2}\sum_{\sigma\neq \sigma^\prime} \sum_{i,j=1 \sim 3(i \neq j)} n_{i\sigma}n_{j\sigma^{\prime}}  \\
 &&+\frac{J_d}{2} \sum_{i,j=1 \sim 3(i \neq j)}( d_{i\uparrow}^{\dag}d_{j\uparrow }d_{j\downarrow }^{\dag}d_{i\downarrow } + {\rm h.c.}), 
\end{eqnarray}
where $\Delta_{i}^d$'s are the one-body levels of the d orbitals as in Fig. \ref{Fig4a}.  
$U_d$, $U_{d}^{\prime}$ and $J_{d}$ represent the intra- and inter orbital Coulomb interaction, and the FM exchange interaction corresponding to the Hund's rule coupling. 
Here, we set these parameters as $U_d=5.0t_{dN}$, and $\Delta_{i}^d = \Delta^d$ ($i=1 \sim 3$): i.e., the CEF and the spin-orbit coupling are neglected because  the energy scale of the CEF and the spin-orbit coupling is much smaller than the Coulomb interaction between d orbitals. 
We will consider the effect of these terms to discuss the anisotropy of the $\pi$- d interaction in \S 4 and \S 5.

Finally, $H_{Pc}$ in eq. (\ref{model_Mec}) is given as
\begin{eqnarray}
H_{Pc} &=& \sum_{j=1 \sim 3}\Delta_{j}^M n_{j\sigma}
  - \sum_{j=1,2,5,6,\sigma} \frac{J_{j}^{M}}{2} n_{j\sigma}n_{3\sigma} \\ 
      & & +\sum_{j=1,2,5,6} \frac{J_{j}^{M}}{2} ( \Phi_{j\uparrow}^{\dag} \Phi_{3\uparrow}\Phi_{3\downarrow }^{\dag} \Phi_{j\downarrow } + {\rm h.c.}), 
\end{eqnarray}
where $\Delta_{j}^M$ are the one-body levels of $\Phi_{1}$, $\Phi_{2}$, and $\Phi_{3}$ measured from $\Phi_{5}$ or $\Phi_{6}$, respectively, for Pc.
The level scheme of $\Delta_{j}^M$'s is shown in Fig. \ref{Fig4a} where we take $\Delta_{1}^M=\Delta_{2}^M \equiv \Delta_{L+1} =3.5$eV and $\Delta_{3}^M \equiv \Delta_{L} =2.0$eV because $\Delta_{L+1} \simeq 1.18t \sim 3.54$eV and $\Delta_{L} \simeq 0.68\sim 2.04$eV for $t \simeq 3$eV as shown in Fig. \ref{Fig1}. 
$J_{j}^{M}$ represents the FM exchange interaction between LUMO and $j$-th $\pi$ orbital.
By using the wave functions obtained in the $\pi$-electron approximation, the magnitude of $J_{1}^M=J_{2}^M \equiv J_L$ and $J_{4}^M=J_{5}^M \equiv J_H$ are evaluated as 
\begin{eqnarray}
J_L &=& \int\int {\rm d}{\bf r}_1{\rm d}{\bf r}_2 \Phi_{1}^{*}({\bf r}_1)\Phi_{3}^{*}({\bf r}_2)\frac{e^2}{|{\bf r}_1 - {\bf r}_2|} \Phi_{1}({\bf r}_2)\Phi_{3}({\bf r}_1), \nonumber \\
    &\simeq& \sum_{j}|c_{j}^{\Phi_{1}}|^2|c_{j}^{\Phi_{3}}|^2 U_p \simeq  0.04U_{p}, 
\end{eqnarray}
and, 
\begin{eqnarray}
J_H \simeq \sum_{j}|c_{j}^{\Phi_{3}}|^2|c_{j}^{\Phi_{4}}|^2 U_p \simeq 0.01U_p,
\end{eqnarray}
where the coefficient, $c_{j}^i$, is the amplitude of the wave function of j-th N and C atoms of $i$-th $\pi$ orbital in eq.(1), and we have approximated as $U_N =U_C = U_p$ where $U_N$ and $U_C$ are intra-Coulomb interactions of $\pi$ orbital on N sites and on C sites. 
We have neglected other interactions. 
Then, we find that $J_L \simeq 4J_H$.

Figure \ref{Fig4a} shows a schematic picture of the Hamiltonian, eq. (\ref{model_Mec}).
Since it is assumed that Fe is trivalent (Fe$^{3+}$), and Pc is monovalent (Pc$^{1-}$) in this Hamiltonian, total number of electrons is 5(d orbitals) + 1(LUMO) + 4 (HOMO+1,HOMO+2) = 10.

\begin{figure}[h]
\begin{center}
\rotatebox{0}{\includegraphics[angle=0,width=1\linewidth]{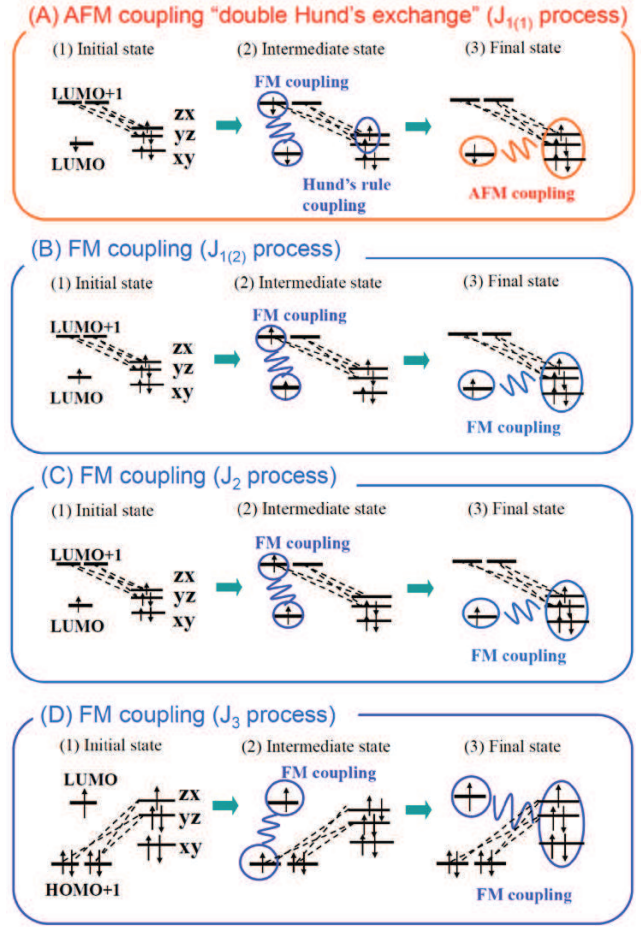}}
\caption{(Color online) Second order perturbation processes of (A) an antiferromagnetic (AFM) interaction ($J_{1(1)}$ process) and (B) $\sim$ (D)  a ferromagnetic (FM) interaction ($J_{1(2)} \sim J_3$ processes).  
The dotted lines indicate the hybridization between d orbital and $\pi$ orbital.}
\label{Fig4b}
\end{center}
\end{figure}

We derive the $\pi$-d interaction in the Hamiltonian eq. (\ref{model_Mec}) by the second order perturbation as follows.
\begin{eqnarray}
H_{\pi-d} &=& 2J_{\pi-d}{\bf S}^{L} \cdot ({\bf S}^{yz} +{\bf S}^{zx}) +2J_{\pi-d}^{\prime}{\bf S}^{L} \cdot {\bf S}^{xy},  \label{pid}
\end{eqnarray}
where ${\bf S}^{L}$, ${\bf S}^{yz}$, ${\bf S}^{zx}$, and ${\bf S}^{xy}$ are the spin operators of LUMO, $d_{yz}$, $d_{zx}$, and $d_{xy}$ orbitals, respectively.
The exchange interaction $J_{\pi-d}$ and $J_{\pi-d}^{\prime}$ contain various processes:
$J_{\pi-d} = J_{1(1)} +J_{1(2)} +J_{2} +J_{3}$, and $J_{\pi-d}^{\prime} = 2(J_{1(1)}+J_{1(2)})$.
These processes are shown in Fig. \ref{Fig4b}(A) $\sim$ Fig. \ref{Fig4b}(D), and the corresponding exchange interactions are given as   
\begin{eqnarray}
J_{1(1)} &=& \frac{J_{L}(t_{11}^2 +t_{12}^2)}{(\Delta_{L+1} -\Delta_{d} -U_d-3U^{\prime}_d  +J_{d})^2 -J_{L}^2}, \\ \label{J1(1)}   
J_{1(2)} &=& -\frac{J_{L}(t_{11}^2 + t_{12}^2)}{(\Delta_{L+1} -\Delta_{d} -U_d -3U^{\prime}_d +2J_d)^2 -J_{L}^2}, \\ \label{J1(2)} 
J_{2} &=&-\frac{2J_{L}(t_{11}^2 + t_{12}^2)}{(\Delta_{L+1} -\Delta_{d} -4U^{\prime}_d +2J_d )^2 -J_{L}^2}, \\ \label{J2} 
J_{3} &=& -\frac{2J_{H}(t_{15}^2 +t_{16}^2)}{(\Delta_{d}+U_d+4U^{\prime}_d -2J_d)^2-J_{H}^2}. \label{J3}
\end{eqnarray}
Note that $J_{1(1)}$ is AFM, while $J_{1(2)}$, $J_{2}$, and $J_{3}$ are FM.
In the following, we explain the second order perturbation processes in detail for $J_{1(1)}$, $J_{1(2)}$, $J_{2}$, and $J_{3}$, respectively.

In Fig. \ref{Fig4b}, the initial state is (xy)$^{2}$(yx)$^{2}$(zx)$^{1}$ for d electrons. 
At first, the schematic picture in Fig. \ref{Fig4b}(A) (the $J_{1(1)}$ process ) corresponds to the following process:
  \begin{description}
\item[(1)] Initial state: (xy)$^2$(yz)$^2$(zx)$^1$.
\item[(2)] Intermediate state: The electron with down spin in the d$_{yz}$ orbital moves to LUMO+1. Because of the Hund's rule coupling between d$_{yz}$ and d$_{zx}$, this intermediate state is favorable. 
On the other hand, if LUMO has a down spin electron, this intermediate state is also favorable because there is a FM exchange coupling between LUMO and LUMO+1.
In this process, we have an energy denominator $(\Delta_{L+1} -\Delta_{d} -U_d-3U^{\prime}_d  +J_{d})^2 -J_{L}^2$ as shown in eq. (\ref{J1(1)}).
\item[(3)] Final state: The electron on LUMO+1 moves back to the d$_{yz}$ orbital.
As a result, we find that the interaction between LUMO and $d_{zx}$ is AFM.  
\end{description}
Secondly the schematic picture in Fig. \ref{Fig4b}(B) (the $J_{1(2)}$ process ) is 
  \begin{description}
\item[(1)] Initial state: (xy)$^2$(yz)$^2$(zx)$^1$.
\item[(2)] Intermediate state: The electron with up spin in the d$_{yz}$ orbital moves to LUMO+1. 
Although there is no energy gain of the Hund's rule coupling between d$_{yz}$ and d$_{zx}$, there exists the FM exchange coupling between LUMO and LUMO+1.
Then, we have an energy denominator $(\Delta_{L+1} -\Delta_{d} -U_d -3U^{\prime}_d +2J_d)^2 -J_{L}^2$ as shown in eq. (\ref{J1(2)}).
\item[(3)] Final state: The electron on LUMO+1 moves back to the d$_{yz}$ orbital.
As a result, we find that the interaction between LUMO and $d_{zx}$ is FM.  
\end{description}

Thirdly, the schematic picture in Fig. \ref{Fig4b}(C) (the $J_{2}$ process ) is 
  \begin{description}
\item[(1)] Initial state: (xy)$^2$(yz)$^2$(zx)$^1$.
\item[(2)] Intermediate state: The electron with up spin in the d$_{zx}$ orbital moves to LUMO+1. 
Although there is no energy gain of the Hund's rule coupling between d$_{yz}$ and d$_{zx}$ as in the case of (B), there exists the FM exchange coupling between LUMO and LUMO+1.
In this case, the $d_{yz}$ orbital is the double occupancy.
Then we have an energy denominator $(\Delta_{L+1} -\Delta_{d} -4U^{\prime}_d +2J_d )^2 -J_{L}^2$.
\item[(3)] Final state: The electron on LUMO+1 moves back to the d$_{zx}$ orbital.
As a result, we find that the interaction between LUMO and $d_{zx}$ is FM.  
\end{description}
Finally, the schematic picture shown in Fig. \ref{Fig4b}(D) (the $J_{3}$ process ) is 
\begin{description}
\item[(1)] Initial state: (xy)$^2$(yz)$^2$(zx)$^1$.
\item[(2)] Intermediate state: The electron with down spin in HOMO+1 moves to the $d_{zx}$.
Because of the FM exchange coupling between $\pi$ orbitals, the energy denominator is $(\Delta_{d}+U_d+4U^{\prime}_d -2J_d)^2-J_{H}^2$.
\item[(3)] Final state: The electron in the $d_{zx}$ orbital moves back to HOMO+1.
Consequently, we find that the interaction between LUMO and $d_{zx}$ is FM.  
\end{description}
From these results, it is found that the process of Fig. \ref{Fig4b}(A) is AFM, while the processes of Fig. \ref{Fig4b}(B) $\sim$ Fig. \ref{Fig4b}(D) are FM.

These mechanisms for exchange interaction are basically the Kanamori-Goodenough (KG) mechanism, because they are between two orthogonal orbitals and they use ferromagnetic exchange interactions between $\pi$ orbitals in the intermediate state~\cite{Kanamori}.
However, they are  much more complicated than the conventional KG mechanism discussed in the TM-oxides, and they can be called  as "generalized" KG mechanisms.
In particular, the process (A) uses two kinds of Hund's rule couplings both on the $\pi$-orbital and on the Fe d-orbital.
This process leads to the antiferromagnetic coupling $J_{1(1)}$, in contrast to the ferromagnetic one in the usual KG mechanism.
Thus, we call this process (A) as a "double-Hund's exchange".
Since this process is favorable for both of the Hund's rule coupling, the energy denominator is generally small among the processes in Fig. \ref{Fig4b}(A) $\sim$ Fig. \ref{Fig4b}(C).
As a result, this "double Hund's exchange" (the process of Fig. \ref{Fig4b}(A)) can be dominant in in Fig. \ref{Fig4b}(A) $\sim$ Fig. \ref{Fig4b}(C).
Although it has been believed that the $\pi$-d interaction is caused by a simple super-exchange type interaction, we found that the mechanism of $\pi$-d interaction in Pc is a new kind of mechanism because of the absence of LUMO-d hopping.

The effective Hamiltonian in  the second order perturbations for (xy)$^{2}$(yx)$^{2}$(zx)$^{1}$ and (xy)$^{2}$(yx)$^{1}$(zx)$^{2}$, are the same, because the amplitude of the transfer integrals between $d_{zx}$ and $\pi$ orbitals are the same as those between $d_{yz}$ and $\pi$ orbitals. 
On the other hand, when the initial state is (xy)$^1$(yz)$^2$(zx)$^2$, there exists only the process as in (A) and (B) of Fig. \ref{Fig4b}.
Then, the exchange interaction is given by $J_{\pi-d}^{\prime}$.

\begin{figure}[h]
\begin{center}
\rotatebox{0}{\includegraphics[angle=0,width=0.8\linewidth]{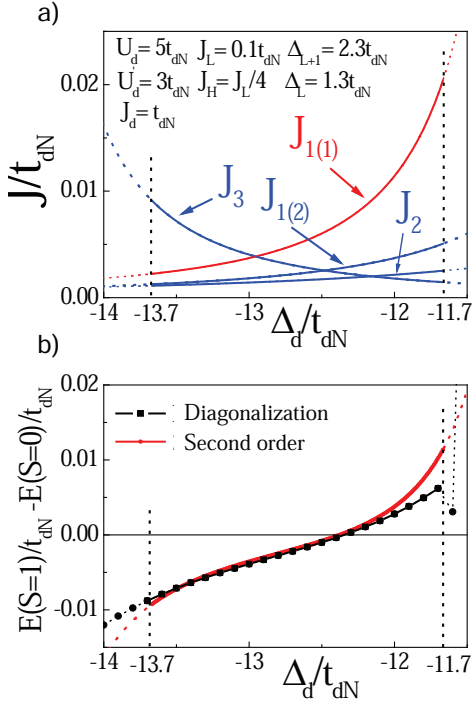}}
\caption{ (Color online) (a) $\Delta_d/t_{(a) {\rm{dN}}}$ dependence on $J_{1(1)}$, $J_{1(2)}$, $J_{2}$, and $J_{3}$ at $U_{d}^{\prime}/t =3.0$, $J_{d}/t =1.0$. 
The blue (red) lines indicate a ferromagnetic (an antiferromagnetic) interaction.
The realistic parameter region of $\Delta_{d}/t_{dN}$ is  $-13.7 < \Delta_{d}/t_{dN} < -11.7$. (b)Energy difference between $S=0$ and $S=1$ states as a function of $\Delta_{d}/t_{dN}$. The black dash-dotted line and the red solid line indicate the results of numerical diagonalization and second order perturbation, respectively.
}
\label{Fig5}
\end{center}
\end{figure}

Figure \ref{Fig5}(a) shows the absolute values of $J_{1(1)}$, $J_{1(2)}$, $J_{2}$, and $J_{3}$ (eqs.(\ref{J1(1)}) $\sim$ (\ref{J3})) as a function of $\Delta_d/t_{{\rm{dN}}}$ for a parameter set $U_{d}^{\prime}/t_{{\rm{dN}}} =5.0$, $U_{d}^{\prime}/t_{{\rm{dN}}} =3.0$, $J_{d}/t_{{\rm{dN}}} =1.0$ and $J_L/t_{{\rm{dN}}} =0.1$.
It is noted that magnitude of various matrix elements associated with Coulomb interactions may be estimated quantitatively by the quantum chemical calculations. However, since these parameters of Fe-Pc system has not been studied in detail, we assume a range of parameter values, which we think reasonable based on the present understanding of transition-metal oxides.
We can see that $J_{1(1)}$ is the largest among these exchange interactions for $\Delta_d/t_{{\rm{dN}}} \gsim -13$, while $J_{3}$ is the largest for $\Delta_d/t_{{\rm{dN}}} \lsim -13$.
It is noted that there is a restriction of the value of $\Delta_{d}/t_{dN}$ as shown below:
we estimate the total energies of (3d)$^4$, (3d)$^5$, and (3d)$^6$ state in the zeroth-order Hamiltonian of $H_{pc} +H_{d}$.  
When $\Delta_{d}/t_{dN} \lsim -13.7$, the electronic state (3d)$^6$ becomes more stable than (3d)$^5$, which is inconsistent with the experiments.
Similarly, when $\Delta_{d}/t_{dN} \gsim -11.7$, (3d)$^4$ becomes more stable than (3d)$^5$, inconsistent with the experiments.
Therefore, the parameter region of the (3d)$^5$ state is $-13.7 \lsim \Delta_{d}/t_{dN} \lsim -11.7$ which is shown in Fig. \ref{Fig5}(a).

In order to check the validity of the second-order perturbation, we evaluate the energy difference between $S=0$ and $S=1$ by a numerical diagonalization of the total Hamiltonian $H =H_{t} + H_{d} + H_{Pc}$.
Figure \ref{Fig5}(b) shows the result (black dashed-dotted line) as a function of $\Delta_{d}/t_{dN}$.
One finds that the ground state is $S=1$ state for $-13.7 \lsim \Delta_d/t_{\rm{dN}} \lsim -12.5$, and the interaction is FM, while the ground state is $S=0$ state for $-12.5 \lsim \Delta_d/t_{dN} \lsim -11.7$, and the interaction is AFM. This indicates that the interaction between LUMO and $d_{yz}$ ($d_{zx}$) orbital can be FM or AFM depending on the parameters.
For comparison, the energy difference between $S=0$ and $S=1$ is also calculated in the second-order perturbation  by $J_{\pi-d}$.
This is shown in Fig. \ref{Fig5}(b) by a red line.
It is found that the result of perturbation is consistent with the result of the numerical diagonalization. 
The reason why these results are in good agreement with each other is that many body effects in d electrons is much stronger than the transfer integrals.
We will discuss and determine the sign of the exchange interaction by comparing with experimental results in \S 6.

\section{Electronic state of d electrons: Analysis of the g-values}
In the previous section, we have discussed the mechanism of $\pi$-d interaction.In order to make a detailed comparison with experiments, it is necessary to take account of the anisotropy of $\pi$-d interaction observed experimentally. 
This anisotropy is due to the spin-orbit interaction and the CEF in the real system.
For the discussion of the anisotropy, we consider a material, PNP[FePc(CN)$_2$], which has a local moment $S=1/2$ of Fe ion, while it does not have $\pi$ electrons (i.e, band insulator).
In this material, we can understand the electronic state purely of d electrons on Fe site.  
Thus, in this section we discuss the electronic state of $d$ electrons for this PNP-compound including its anisotropy, by analyzing the anisotropy of the g-values obtained in ESR experiment~\cite{Hanasaki1}.  

\begin{figure}[h]
\begin{center}
\rotatebox{0}{\includegraphics[angle=-90,width=1\linewidth]{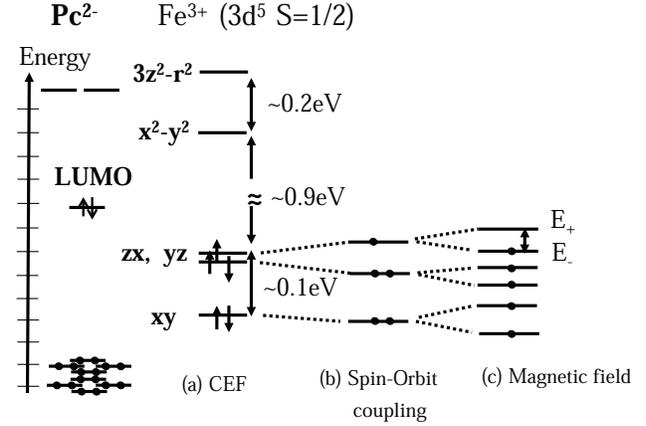}}
\caption{Schematic picture of the energy levels in t$_{2g}$ orbitals with (a)the crystalline electric field (CEF) effect, (b)the spin-orbit coupling , and (c) the magnetic field,  comparing with Pc$^{2-}$ electric structure.
}
\label{Fig2}
\end{center}
\end{figure}

Figure \ref{Fig2}(a) shows the schematic picture of the CEF splitting of the d orbitals of PNP[FePc(CN)$_2$], where the energy splittings are estimated on the basis of the usual CEF theory~\cite{Sugano}.
Since it has been known that Fe$^{3+}$ is in a low spin state from the static magnetic susceptibility measurements, five electrons occupy $t_{2g}$ ($d_{xy}$, $d_{yz}$, and $d_{zx}$) orbitals as shown in Fig. \ref{Fig2}(a).

\begin{figure}[h]
\begin{center}
\rotatebox{0}{\includegraphics[angle=-90,width=1\linewidth]{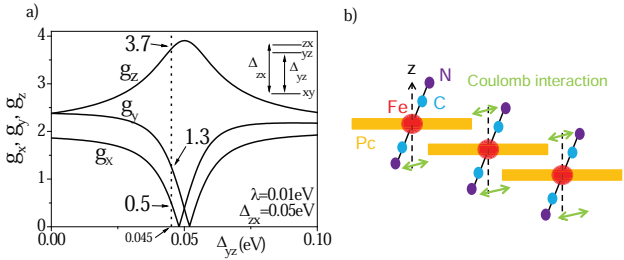}}
\caption{(Color online) (a) $\Delta_{yz}$ dependence on $g_x$, $g_y$ and $g_z$ at $\Delta_{zx}=0.05$eV and $\lambda=0.01$eV.
The dotted line indicates the location of $\Delta_{yz}=0.045$eV.
(b) Schematic picture of the stacking of FePc(CN)$_2$ unit.
 }
\label{Fig3}
\end{center}
\end{figure}

Since the energy difference between $t_{2g}$ and $e_g$ states is about 1eV, we consider only $t_{2g}$ orbitals to discuss the g-values. 
Then, an effective Hamiltonian becomes 
\begin{eqnarray}
H_{{\rm tot}} =  H_{{\rm 0}} + H_{{\rm so}} + H_{{\rm{mag}}}^{p},
\label{Hamiltonian} 
\end{eqnarray}  
where $H_{{\rm 0}}$, $H_{{\rm so}}$, and $H_{{\rm{mag}}}^{p}$ represent one-body energy level, spin-orbit coupling, and the magnetic field along the $p$ direction ($p=x,y,z$), respectively.
We write $H_{{\rm 0}}$ as 
\begin{eqnarray}
H_{0} =\sum_{\sigma}[\Delta_{yz}n_{yz\sigma} +\Delta_{zx}n_{zx\sigma}], \label{ESD}
\end{eqnarray}  
where $\Delta_{yz}$ and $\Delta_{zx}$ are one-body levels of $d_{yz}$ and $d_{zx}$ measured from $d_{xy}$, respectively.
Note that the amplitude of $\Delta_{yz}$ is generally different from that of $\Delta_{zx}$ because of the distortion in the local structure around Fe (See Fig. \ref{Fig3}(b)).    
The spin-orbit coupling in the manifold of $t_{2g}$ orbitals is given as~\cite{Ng}
\begin{equation}
H_{{\rm so}}={\rm i}{\lambda\over 2}\sum_{\ell,m.n}\epsilon_{\ell mn}
\sum_{\sigma,\sigma'}d^{\dagger}_{\ell\sigma}d_{m\sigma'}
(\sigma_{n})_{\sigma\sigma'},  \label{SO}
\end{equation}
where $\sigma_{n}$ is the $n$-th component of the Pauli matrix, i.e. $\sigma_{1}=\sigma_{x}$, $\sigma_{2}=\sigma_{y}$, and $\sigma_{3}=\sigma_{z}$.
$\lambda$ is a spin-orbit coupling constant, and $\epsilon_{\ell mn}$ is the 
Levi-Civita symbol. 
$d_{\ell\sigma}$ is the annihilation operator with spin $\sigma$ on the $\ell$ orbital.
Finally, The Hamiltonian of the magnetic field is 
\begin{eqnarray}
H_{{\rm{mag}}}^{p} = H_{{\rm{mag}}}^{s_p} + H_{{\rm{mag}}}^{l_p}, 
\end{eqnarray}
where $p=1 \sim 3$ ($1=x$, $2=y$, and $3=z$).
$H_{\rm{mag}}^{s_p}$ and $H_{\rm{mag}}^{l_p}$ are 
\begin{eqnarray}
H_{\rm{mag}}^{s_p} &=&\frac{g_0}{2}h_p\sum_{l (\neq p).\sigma\,\sigma^{\prime}}d_{l\sigma}^{\dag}d_{l\sigma^{\prime}}(\sigma_p)_{\sigma\sigma^{\prime}}, \\
H_{\rm{mag}}^{l_p} &=& {\rm i}h_p\sum_{\ell,m (\neq p)}\epsilon_{\ell m p}
\sum_{\sigma}d^{\dagger}_{\ell\sigma}d_{m\sigma},
\end{eqnarray}
for the $t_{2g}$ orbitals, where $g_0=2$, $h_p =\mu_B H_p$, and $\mu_B$ and $H_p$ is the Bohr magneton, and the external magnetic field along the $p$ axis, respectively.

When five electrons occupy the $t_{2g}$ orbitals, there are six configurations.
On the basis of these configurations, the eigenvalues and eigenstates of the Hamiltonian eq.(\ref{Hamiltonian}) are obtained by a simple diagonalization.
As shown in Fig. \ref{Fig2}(b), these eigenvalues consist of three two-fold degenerate states (Kramers doublets) without magnetic field.
Under the magnetic field, the Kramers doublet splits as shown in Fig. \ref{Fig2}(c), where $E_{+}$ and $E_{-}$ are the highest energy state and the second highest energy state, respectively.
Then, the g-values along the $p$ ($p=1 \sim 3$) axis are given by $g_p = \partial[E_+ -E_-]/\partial h_p \bigr|_{h_p \rightarrow 0}$.

Figure\ \ref{Fig3}(a) shows the numerical results of $g_p$ ($p=1\sim3$) as a function of $\Delta_{yz}$ with $\Delta_{zx}$ being fixed at $\Delta_{zx}=0.05$ eV and $\lambda$ = 0.01eV.
At $\Delta_{yz}= 0.045$eV (the dotted line in Fig.\ \ref{Fig3}(a)), the g-values are evaluated as $g_x \simeq 0.5 $, $g_y \simeq 1.3$, and $g_z \simeq 3.7$, respectively.
These values are in good agreement with the g-values of PNP[FePc(CN)$_2$], $g_x \simeq 0.5$,$g_y \simeq 1.1$, and $g_z \simeq 3.6$, obtained experimentally~\cite{Hanasaki1}.
As shown in Fig. \ref{Fig3}(b), the difference between $\Delta_{yz}$ and $\Delta_{zx}$ will be due to the distortion of the apical cyanides caused by the Coulomb interaction between the cyanide and the nearest Pc.

\section{Effective model of the $\pi$-d interaction}
Although the mechanism of $\pi$-d interaction is clarified in \S 3, the sign of the $\pi$-d interaction has not been determined.
In this section, we discuss its sign by calculating the susceptibility of a single Fe-Pc system which has one localized spin on Fe and on the LUMO, respectively.
We use an effective Hamiltonian  
\begin{eqnarray}
H_{\pi-d}^{tot} =  H_{0} +H_{so} +H_{\pi-d},  \label{pdtot}
\end{eqnarray}
where $H_{0}$, $H_{so}$, and $H_{\pi-d}$ are eq. (\ref{ESD}), eq. (\ref{SO}), and eq. (\ref{pid}), respectively. 
Note that, in deriving $H_{\pi-d}$, we have neglected the effects of CEF and spin-orbit coupling assuming that these effects gives subsidiary modifications of exchange interaction.
In eq. (\ref{pdtot}), we discuss the $\pi$-d exchange interaction in the presence of $H_{0}$ and $H_{so}$,  in order to make a comparison with experiment.

\begin{figure}[h]
\begin{center}
\rotatebox{0}{\includegraphics[angle=-90,width=1\linewidth]{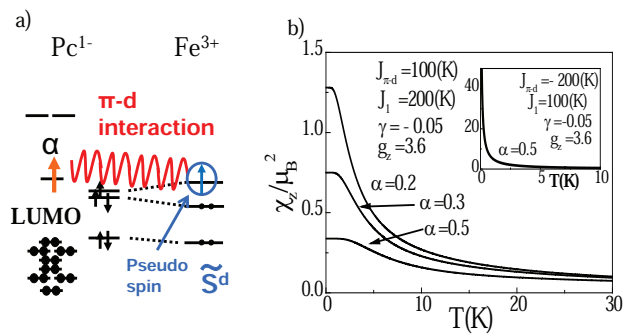}}
\caption{(Color online) (a) Schematic picture of the $\pi$-d interaction between the pseudo spin on LUMO ($\tilde{S}^{L}$) and the pseudo spin on d orbitals ($\tilde{S}^{d}$).  $\alpha$ is the parameter reflecting the electronic state of the conduction band. (b) Temperature dependence of the susceptibility in the case of the antiferromagnetic $\pi$-d interaction. Inset: Temperature dependence of the susceptibility in the case of the ferromagnetic $\pi$-d interaction.
}
\label{Fig6}
\end{center}
\end{figure}

As shown in Fig. \ref{Fig6}(a), the ground state of d electrons with the spin-orbit coupling and CEF effect consist of three Kramers doublets (Same as Fig. \ref{Fig2}(b)).
We express the electron in the highest Kramers doublet as a pseudo spin~\cite{Lines}.
We introduce the pseudo spin operators as ${S}_{x}^d \equiv \gamma \tilde{S}_{x}^d$, ${S}_{y}^d \equiv \gamma \tilde{S}_{y}^d$, and ${S}_{z}^d \equiv \gamma^{\prime} \tilde{S}_{z}^d$, where ${\bf S}^d \equiv {\bf S}^{yz} + {\bf S}^{zx} +{\bf S}^{xy}$, $\gamma \sim -(\lambda/\Delta_{zx})^2$, and $\gamma^{\prime} \sim 1$.
It is noted that the base of pseudo spin $\tilde{\bf S}^d$ is the highest Kramers doublet.
We also find $S_{x}^{xy}  \simeq \gamma \tilde{S}_{x}^d$, $S_{y}^{xy}  \simeq \gamma \tilde{S}_{y}^d$, and $S_{z}^{xy} = \gamma^{\prime \prime} \tilde{S}_{z}^d$ where  $\gamma^{\prime \prime} \sim (\lambda/\Delta_{zx})^2$.
The absolute values of $\gamma$ and $\gamma^{\prime\prime}$ are of the order of $0.01$ for $\Delta_{zx} =0.05$eV and $\lambda =0.01$eV.  Note that LUMO is the part of the conduction band constructed from $\pi$ orbitals in actual materials. 
Therefore, we also introduce a pseudo-spin operators of LUMO as ${\bf S}^{L} \equiv \alpha\tilde{\bf S}^{L}$ where $\alpha$ is a parameter reflecting the electronic state of the conduction band.

By using these pseudo spins, the effective Hamiltonian is given by
 \begin{eqnarray}
H_{\pi-d}^{eff} \simeq 2\tilde{J}_{1}\gamma (\tilde{S}^{L}_{x}\tilde{S}_{x}^d +\tilde{S}^{L}_{y}\tilde{S}_{y}^d ) +2\tilde{J}_{\pi-d}\tilde{S}^{L}_z \tilde{S}_z^d,
\end{eqnarray}
where $\tilde{J}_{1}= \alpha J_{1} \equiv \alpha(J_{1(1)} +J_{1(2)})$, $\tilde{J}_{\pi-d}= \alpha J_{\pi-d}$, and we have set as $\gamma^{\prime} + \gamma^{\prime\prime}=1$.
The red wiggly line of Fig. \ref{Fig6}(a) indicates the interaction between $\tilde{\bf S}^{L}$ and $\tilde{\bf S}^{d}$ ($\pi$-d interaction). 

The effective Hamiltonian of the magnetic field is given by 
 \begin{eqnarray}
H_{mag}^z= [g_{z}^{L}\mu_B\alpha\tilde{S}_{z}^{L} + g_{z}^{d}\mu_B\tilde{S}_{z}^{d}]h_{z},
  \end{eqnarray}
 where $h_{z}$ is the magnetic field along z axis, and $g_{z}^{L}$ and $g_{z}^{d}$ are the g-values of pseudo spin on the $\pi$ orbitals and d orbitals, respectively.

Figure \ref{Fig6}(b) shows the temperature $T$-dependence of the susceptibility by a simple diagonalization for a parameter set $J_{\pi-d} =100K$ (the AFM interaction), $J_{1}=200K$, $\gamma =-0.05$ and $g_{z}^L=2.0$ with $\alpha =0.2$, $0.3$, and $0.5$.
Here, we set g factor as $g_{x}^d \sim 0.5$, $g_{y}^d \sim 1.1$, and $g_{z}^d \sim 3.6$ from the experimental result of PNP[FePc(CN)$_2$]$_2$.
For $T \gsim 5K$, the $T$-dependence is Curie's law, while for $T \lsim 5K$ the susceptibility is constant irrespective of $\alpha$ due to the antiferromagnetic coupling.  
The inset of Fig \ \ref{Fig6}(b) shows the $T$-dependence for a parameter set $J_{\pi-d} =-200K$ (the FM interaction), $J_{1} =100K$, $\gamma =-0.05$, $g_{z}^d=3.6$ and $g_{z}^L=2.0$ with $\alpha =0.5$.
The amplitude of FM interaction is taken as the same order of the result of Ref~\cite{Yu}. 
The $T$-dependence is Curie's law over the whole temperature range, because the ground state is a doublet of $(\tilde{S}^{t},\tilde{S}^{t}_z) =(1,\pm 1)$ where  $\tilde{S}^{t}$ and $\tilde{S}^{t}_z$ are a total pseudo spin and z component of the pseudo spin.   

Here, let us compare the above theoretical result with experiments.
TPP [Fe$_{0.07}$Co$_{0.93}$Pc(CN)$_2$]$_2$ is a suitable system for comparison, because it contains dilute Fe local spins.
It is reported that the susceptibility of TPP [Fe$_{0.07}$Co$_{0.93}$Pc(CN)$_2$]$_2$ increases as $T$ decreases, and then it decreases in the range  $T < 10{\rm K}$~\cite{Hanasaki4}. Since the $T$-dependence of the susceptibility of TPP[CoPc(CN)$_2$]$_2$ is small and almost constant except for the low temperature~\cite{Hanasaki3}, the origin of the $T$-dependence of TPP[Fe$_{0.07}$Co$_{0.93}$Pc(CN)$_2$]$_2$ must be due to the Fe local spin with the $\pi$-d interaction.
By comparing this experimental result with the theoretical results shown in Fig.\ref{Fig6}(b), it is found that the exchange interaction should be AFM, because the susceptibility does not diverge at the low temperatures.
For $\alpha \sim 0.3$, the $T$-dependence in Fig. \ref{Fig6}(b) is similar to the experimental result. 

When we compare the experimental results more closely, we find that $\chi$ decreases slightly below $T \sim10$ K.
This behavior is not obtained theoretically in Fig. \ref{Fig6}(b).
We expect that this decrease of $\chi$ is due to the Kondo effect which has not been taken into account in our model.
In order to take account of the Kondo effect, the Fermi surface of the $\pi$-band should be included, which remains as a future problem.
In this case, the charge disproportionation ($4k_F$ CDW) observed in TPP[CoPc(CN)$_2$]$_2$ should be also taken into account~\cite{Hanasaki3}.
Although these considerations are necessary for the detailed comparison, the conclusion of the AFM exchange interaction between d electrons and $\pi$ electrons will not be changed.

As discussed above, we found that the $\pi$-d interaction is AFM from the experimental result, while it has been suggested that the $\pi$-d interaction is FM in the quantum chemical calculation~\cite{Yu}.
Since the AFM interaction occurs by the second-order perturbation using the high energy states (LUMO+1 and LUMO+2), we suspect that these excited states are not taken into account enough in the quantum chemical calculation.
The extended Kondo lattice model assuming FM or AFM $\pi$-d interaction has been suggested to be the origin of the negative magnetoresistance~\cite{Hotta1,Otsuka,Hotta2}.
From our study, one finds that the AFM Kondo lattice model suggested by ref. \cite{Hotta1} is in agreement with our result.
However, because an isotropic $\pi$-d interaction is assumed in ref. \cite{Hotta1}, it is needed that the anisotropy of d electrons in TM is considered.

\section{Conclusion}
In this paper, we have clarified that the mechanism of $\pi$-d interaction in Pc is an extended version of the Kanamori-Goodenough mechanism including the double-Hund's exchange mechanism that is antiferromagnetic.
This is different from the super-exchange mechanism occurring from the hybridization between two orbitals. 
We found that the exchange interaction between d orbitals of Fe and LUMO in Pc can be either the FM or AFM interaction depending on the parameters on the basis of the second order perturbation and numerical diagonalization calculations.
By the comparison with the theoretical result of susceptibility on the basis of the effective model with the anisotropy of exchange interactions, we concluded that the exchange interaction is AFM.

\section{Acknowledgements}
This work was partly supported by a Grant-in-Aid for Scientific Research on Priority Areas of Molecular Conductors (No. 15073210) from the Ministry of Education, Culture, Sports, Science and Technology, Japan (MEXT),and also by a Next Generation Supercomputing Project, Nanoscience Program, MEXT, Japan.


\end{document}